\documentclass[11pt]{article} 
\usepackage{epsfig}
\usepackage{natbib}
\begin{document}

\newcommand{\be}{\begin{equation}}
\newcommand{\ee}{\end{equation}}
\newcommand{\bea}{\begin{eqnarray}}
\newcommand{\eea}{\end{eqnarray}}
\newcommand{\da}{\dagger}
\newcommand{\dg}[1]{\mbox{${#1}^{\dagger}$}}
\newcommand{\hlf}{\mbox{$1\over2$}}
\newcommand{\lfrac}[2]{\mbox{${#1}\over{#2}$}}
\newcommand{\scsz}[1]{\mbox{\scriptsize ${#1}$}}
\newcommand{\tsz}[1]{\mbox{\tiny ${#1}$}}

\begin{flushright} 
\end{flushright} 

\begin{center}

\Large{\bf Lunar Laser Ranging Science: Gravitational Physics and Lunar Interior and Geodesy\rm\footnote{Presented at 35th COSPAR Scientific Assembly, July 18-24, 2004, Paris, France, 2004}}

\vspace{0.4in}

\normalsize
\bigskip 

James G. Williams, Slava G. Turyshev, Dale H. Boggs, and J. Todd Ratcliff

\normalsize
\vskip 15pt

{\it{Jet Propulsion Laboratory, California Institute of  Technology,\\
Pasadena, CA 91109, U.S.A.}} 
\footnote{Email: {\tt James.G.Williams@jpl.nasa.gov, turyshev@jpl.nasa.gov, Dale.H.Boggs@jpl.nasa.gov, \\ J.Todd.Ratcliff@jpl.nasa.gov }}
 

\vspace{0.25in}

\end{center}


\begin{abstract}
Laser pulses fired at retroreflectors on the Moon provide very accurate ranges.  Analysis yields information on Earth, Moon, and orbit.  The highly accurate retroreflector positions have uncertainties less than a meter.  Tides on the Moon show strong dissipation, with $Q=33\pm4$ at a month and a weak dependence on period.  Lunar rotation depends on interior properties; a fluid core is indicated with radius $\sim$20\% that of the Moon.  Tests of relativistic gravity verify the equivalence principle to $\pm1.4\times 10^{-13}$, limit deviations from Einstein's general relativity, and show no rate for the gravitational constant $\dot{G}/G$ with uncertainty $9\times 10^{-13}$ yr$^{-1}$.  

\end{abstract}
\vspace{0.15in}


\section{Introduction}

Lunar Laser Ranging (LLR) accurately measures the time of flight of a laser pulse fired from an observatory on the Earth, bounced off of a corner cube retroreflector on the Moon, and returned to the observatory.  These ranges contain information on the Earth, the Moon, and the orbit.  The first retroreflector array was placed on the lunar surface in 1969 by the Apollo 11 astronauts.  Additional reflectors were left on the Moon by the Apollo 14 and Apollo 15 astronauts, and French-built reflector arrays are on the Lunokhod 1 and 2 rovers transported by the Soviet Luna 17 and Luna 21 missions.  

LLR continues to produce valuable results after 35 yr because a sequence of improvements at the ranging stations have decreased the range uncertainty.  The LLR data are analyzed using a weighted least-squares approach.  After the fits the rms residual is several decimeters in the first decade improving to 2 cm over the last decade.  Currently operating stations are at McDonald Observatory and Observatoire de la Côte d'Azur. 
 
A general review of LLR is given by \cite{Dickey_etal_1994}.  A discussion of recent science results and a comparison of Earth and Moon are given by \cite{Williams_Dickey_2003}.  This paper discusses positions on the Moon, along with lunar rotation, tides and fluid core, and then tests of gravitational physics.  

\section{Positions on the Moon}
\label{sec:pos_moon}
\subsection{Current position capabilities}
  
Operational ranging is done to the retroreflector arrays at the Apollo 11, 14, and 15 sites plus the French-built reflector on the Soviet Lunokhod 2.  The resulting reflector coordinates with respect to the center of mass have submeter uncertainties.  These four sites have the most accurately known positions on the Moon.  

Precise lunar geodesy depends on an accurate knowledge of the time-varying three-dimensional lunar rotation.  The lunar coordinate frame and the Euler angles are interconnected.  The Earth's coordinate frame has the $Z$ axis is near the principal axis of greatest moment of inertia while the direction of zero longitude is adopted.  The synchronous lunar rotation permits all three lunar axes to be defined dynamically.  Two options for the definition are principal axes or axes aligned with the mean Earth and mean rotation directions.  The three-dimensional rotation of the Moon, called physical libration, is integrated numerically along with the lunar and planetary orbits.  For the integration and fits the lunar coordinate system is aligned with the principal axes since that is a natural frame for expressing the torques and differential equations for rotation, integrating the rotation, and fitting the data.  Principal axes are used with the accurate numerically integrated lunar rotation, expressed as Euler angles, which is distributed with modern JPL ephemerides such as DE403 ({\tt http://ssd.jpl.nasa.gov/}).  When using principal axes the gravitational coefficients $C_{21}, S_{21}$, and $S_{22}$ are zero.  

For a synchronously rotating triaxial ellipsoid the principal axes coincide with the mean Earth and mean rotation axes.  For less symmetrical bodies, lunar gravitational harmonics which are even in latitude and odd in longitude, such as $S_{31}$ and $S_{33}$, cause a small constant rotation of the $X$ and $Y$ principal axes about the $Z$ axis so that the $X$ principal axis is displaced in longitude from the mean direction to the Earth.  Similarly, harmonics which are even in longitude and odd in latitude, e.g. $C_{30}$ and $C_{32}$, cause a constant rotation of the $X$ and $Z$ principal axes about the $Y$ axis so that the $Z$ principal axis is displaced from the mean axis of rotation and the $X$ axis is displaced in latitude from the mean Earth direction.  These rotations depend on the harmonics of the ephemeris or solution.  For DE403 the shifts are 64" in longitude and 79" in pole direction so these shifts exceed 1/2 km at the surface for all three axes (1" is 8.43 m at the surface).  A lunar geodetic network can use principal axes or it can elect to remove the constant angular shifts to get the mean Earth and mean rotation directions.  The advantage of the latter choice is that the coordinate values will not be sensitive to changes in the gravity harmonics while high accuracy is the advantage of the former.  A set of retroreflector coordinates is given by \cite{Williams_Newhall_Dickey_1996a} and the coordinates are identified as ``mean Earth/rotation coordinates.''  Later citations have identified these as ``mean Earth/polar axis coordinates,'' but it is essential to understand that the third axis is the mean of the direction of the (Euler-angle-defined) polar rotation axis because this time-varying rotation axis moves $>$1 km with respect to the body.  The angular velocity vector is not coincident with the foregoing rotation axis, it also moves by a substantial amount with respect to the body, and it plays no role in the definition of coordinate frames since it is not used for orientation.  

Compared to uniform rotation and precession the variations of lunar rotation and orientation are up to 1 km.  For best rotation accuracies one should use the numerically integrated physical librations.  If it is wished, the principal axis coordinates can be converted to mean Earth/mean rotation axis coordinates with constant angular shifts, but there is an uncertainty of about 1 m in the longitude shift.  For coarser work an approximate series for orienting the Moon, appropriate for mean Earth and mean rotation axes, is given in \citep{Seidelmann_etal_2002}.  A comparison of the series and numerically integrated physical librations is presented in \citep{Konopliv_etal_2001}. 
 
The LLR reflector coordinates, along with VLBI determined ALSEP transmitter locations \citep{King_Counselman_Shapiro_1976}, serve as control points for a lunar geodetic network \citep{Davies_Colvin_Meyer_1987, Davies_etal_1994}.  Coordinates of features at the Apollo sites are given by \cite{Davies_Colvin_2000}.  Only three of the Apollo sites (15, 16, and 17) are located on the high quality Apollo mapping camera photographs.  The sparse and uneven distribution of accurate control points causes the accuracy of the network to vary strongly with lunar position.  

\subsection{Future opportunities}

Future missions to the Moon could improve the accuracy of the lunar geodetic network in two ways:  a) Future landers should carry accurate radio tracking and, when practical, retroreflectors.  b) Future lunar orbital missions should accurately map all of the Apollo sites plus the Lunokhod sites plus new lander locations.  

Spacecraft positions with a few meters accuracy would be useful for lunar geodesy if the spacecraft were located on high resolution images.  Future landed spacecraft can be positioned with useful accuracy if tracked accurately by radio from the Earth.  As seen from the Earth, the Doppler shift for various locations on the Moon only varies $\sim$2 m/sec so good quality Doppler tracking of $\sim$0.03 mm/sec would give a surface position accuracy of order 10~m for the $Y$ and $Z$ components and worse for the $X$ component.  Range accuracies of 1 to 2 meters would give a good $X$ coordinate accurcy, and $Y$ and $Z$ accuracies of order 5 m if tracked for a month or more.  As seen from the Moon, the Earth oscillates in the sky by $\pm0.1$ radians in both coordinates and this variation gives the most accurate information on the $Y$ and $Z$ coordinates.  Differential radio tracking between sites could also be useful \citep{King_Counselman_Shapiro_1976} since there are common error sources which cancel.  VLBI \citep{Hanada_etal_1999} could give two coordinates well and the third more weakly. 
 
For highest accuracy positions, retroreflectors remain excellent devices to place on the Moon.  Other high accuracy devices, such as optical transponders, may offer future alternatives.  While landing site locations can be recovered with modest data spans, lifetimes of years are needed to improve the lunar rotation and orientation and to extract lunar science.  

\section{Lunar science - a window on the interior}
\subsection{Tides}

The elastic response of the Moon to tidal forces is characterized by Love numbers.  The second-degree tides are strongest and these tidal displacements are sensitive to the second-degree Love numbers $h_2$ and $l_2$ while the rotation is sensitive to the potential Love number $k_2$.  The amplitudes of the two largest monthly terms are both about 9 cm.  In practice, LLR solutions are more sensitive to $k_2$.  A solution, with data from 1970 to 2003, gives $k_2 = 0.0227\pm0.0025$ and $h_2 = 0.039\pm0.010$ \citep{Williams_Boggs_Ratcliff_2004}.  $l_2$ was fixed at a model value of 0.011.  For comparison, there is an orbiting spacecraft value of the lunar Love number $k_2 = 0.026\pm0.003$ determined from tidal variation of the gravity field \citep{Konopliv_etal_2001}.  

For comparison with the solution results, model calculations have been made for lunar Love numbers.  Love number calculations start with an interior model \citep{Kuskov_Kronrod_1998} which is compatible with seismic P- and S-wave speeds deduced from Apollo seismometry.  There is little seismic information below 1100 km and the seismic speeds have to be extrapolated into the deeper regions above the core.  The 350 km radius of the fluid iron core was adjusted to match the LLR-determined $k_2$ and small adjustments were made to the densities to accommodate mass and moment constraints.  In addition to matching $k_2 = 0.0227$, the model calculations give $h_2 = 0.0397$ and $l_2 = 0.0106$.  

The LLR solutions are also sensitive to tidal dissipation \citep{Williams_etal_2001}.  In general, the specific dissipation $Q$ depends on frequency.  In the above solutions, the whole-Moon monthly tidal $Q$ is found to be $33\pm4$.  For $k_2 = 0.0227$ the power-law expression for tidal $Q$ as a function of tidal period is determined to be $33({\rm Period}/27.212{\rm d})0.05$, so the $Q$ increases from 33 at a month to 38 at one year.  At tidal frequencies the Moon exhibits strong dissipation.  

\subsection{Molten core}

Evidence for a distinct lunar core comes from the moment of inertia \cite{Konopliv_etal_1998}, the induced dipole moment \citep{Hood_etal_1999}, and lunar laser ranging.  LLR analyses indicate that the core is fluid.  This is a small dense core, presumably iron rich with elements such as sulfur which lower the melting point.  

In addition to strong tidal dissipation, the lunar rotation displays a strong source of dissipation which is compatible with a fluid core \cite{Williams_etal_2001}.  This source of dissipation arises from the fluid flow along a fluid-core/solid-mantle boundary (CMB).  With the aid of Yoder's (1995) turbulent boundary layer theory these dissipation results give a 1-  upper limit for core radius of 352 km for a pure Fe core or 374 km for a fluid Fe-FeS eutectic \citep{Williams_etal_2001}, about 20\% of the lunar radius.  Upper limits are used because any topography on the CMB or the presence of an inner core would tend to decrease the real radius. 
 
The detection of the oblateness of the fluid-core/solid-mantle boundary (CMB) would be independent evidence for the existence of a liquid core.  Fluid flow along an oblate boundary exerts torques on both the fluid and the overlying solid Moon.  In recent years rotation evidence for an oblate boundary has been strengthening \citep{Williams_Boggs_Ratcliff_2004}.
  
The internal structure and material properties of the Moon must be deduced from external evidence and the deepest regions are the least understood.  In order to determine the variety of viable interior structures and properties, a large number of models have been generated which satisfy, within measurement uncertainties, four lunar quantities: the mean density, the moment of inertia's measure of mass concentration toward the center, the $k_2$ elastic response to solid-body tides, and tidal dissipation $Q$ \citep{Khan_etal_2004}.  Typically, the central regions of the acceptable models have a higher density core which can take several forms such as completely solid, completely fluid, and a solid inner core within a fluid outer core.  

\section{Tests of relativistic gravity}
\subsection{Motivation}

Einstein's general theory of relativity has proved remarkably successful.  Nonetheless, this is a time for tests of gravity in the solar system.  There is an expectation that a theory of gravity can be found which is compatible with the quantum theories of the stronger forces.  Scalar-tensor extensions of gravity that are consistent with present cosmological models predict deviations of relativistic gravity from general relativity.  

Different aspects of metric theories of gravity are described with Parametrized Post-Newtonian (PPN) $\beta$  and $\gamma$   parameters.  These PPN parameters have a unit value for general relativity, but a deviation at levels of $10^{-5}$ to $10^{-7}$ has been predicted by \cite{Damour_Nordtvedt_1993} and further developed by \cite{Damour_Piazza_Veneziano_2002}.  

The great stability of the lunar orbit allows LLR to use the orbital motion to make accurate tests of gravitational physics.  A discussion follows of LLR tests of the equivalence principle, the implication for PPN  $\beta$, and variation of the gravitational constant.  

\subsection{Equivalence principle}

The equivalence principle is a foundation of Einstein's theory of gravity.  The LLR analysis tests the equivalence principle by examining whether the Moon and Earth accelerate alike in the Sun's field.  \cite{Nordtvedt_1968,Nordtvedt_1970} gave theoretical analyses of the effects of a violation of the principle of equivalence.  For the Earth and Moon accelerated by the Sun, if the equivalence principle is violated the lunar orbit will be displaced along the Earth-Sun line, producing a range signature with a 29.53-day period.  

The LLR test of the equivalence principle shows that the Earth and Moon are accelerated alike by the Sun's gravity with ($\Delta $acceleration/acceleration) of $(-1.0\pm1.4)\times10^{-13}$ \citep{Williams_Turyshev_Boggs_2004}.  The uncertainty corresponds to a 4 mm deviation of the lunar distance.  

A violation of the equivalence principle might depend on composition or the strength of the gravitational attraction within a finite body (gravitational self energy).  The former was tested in the laboratory by \cite{Adelberger_2001} with an uncertainty similar to the LLR result.  The latter requires large bodies such as the Moon or planets and depends on PPN  $\beta$ and $\gamma$.  Combining the LLR result, the laboratory composition result, and the Cassini time delay test of   \cite{Bertotti_Iess_Tortora_2003}, one derives  $\beta - 1 = (1.2\pm1.1)\times10^{-4}$ \citep{Williams_Turyshev_Boggs_2004}.  This is the strongest limit on PPN $\beta$  to date and is not significantly different from the unit value of general relativity.  

\subsection{Does the gravitational constant vary?}

Einstein's general theory of relativity does not predict a variable gravitational constant $G$, but some other theories of gravity do.  A changing G would alter the scale and periods of the orbits of the Moon and planets.  LLR is sensitive to $\dot{G}/G$ at the 1 AU scale of the annual orbit about the Sun \citep{Williams_Newhall_Dickey_1996b}.  No variation of the gravitational constant is discernible, with $\dot{G}/G = (4\pm9)\times10^{-13}$ yr$^{-1}$ \citep{Williams_Turyshev_Boggs_2004}.  The uncertainty corresponds to 1.2\% of the inverse age of the universe.  The scale of the solar system does not share the cosmological expansion.  

\section{Future improvements}

There would be scientific benefits if additional retroreflectors were placed on the Moon.  The uncertainty in the derived lunar rotation and the related lunar science parameters depends on the spread of reflector $Y$ and $Z$ coordinates.  These are currently 38\% of the lunar diameter for $Y$ and 25\% for $Z$.  A wider geographic spread would improve sensitivity by several times.  The present configuration, two reflectors near the equator and two at mid-northern latitudes, would be better balanced with one or more southern sites.  Any expansion of the distribution would be beneficial.  

Further improvements can be made in the accuracy of ranging devices.  A highly accurate laser ranging facility is being assembled at the Apache Point Observatory, New Mexico \citep{Murphy_etal_2000} which will have the advantage of a large telescope and modern detectors.  

\section{Summary}

The analysis of LLR data provides information on the Earth, the Moon, and the lunar orbit.  Improvements in range accuracy during the 35 yr data span enable new results to emerge.  This paper discusses LLR results for positions on the Moon, lunar solid-body tides, evidence for a molten lunar core, and tests of equivalence principle and variability of the gravitational constant $G$.  

The four regularly ranged retroreflectors have the most accurately known locations on the Moon with their submeter uncertainties.  Future geodetic control needs additional precise locations with wide distribution.  It is recommended that future lunar surface spacecraft have accurate radio tracking and, when practical, carry retroreflectors to the lunar surface. 
 
The internal structure and material properties of the Moon must be deduced from external evidence and the deepest regions are the most elusive.  LLR provides information on the Moon's tidal response, tidal dissipation, and interactions at the core/mantle interface.  Dissipation at the core/mantle boundary implies that the core is fluid and has a radius about 20\% that of the Moon.  There is still considerable uncertainty about the core's radius and properties, and whether a solid inner core exists, but for the core there is compatibility between model, moment of inertia, rotation, and magnetic evidence. 
 
The equivalence principle is valid with a relative uncertainty of $1.4\times10^{-13}$.  The gravitational constant shows no evidence of variation with $\dot{G}/G = (4\pm9)\times 10^{-13}$ /yr.  LLR tests are in agreement with Einstein's general theory of relativity. 

If future spacecraft carry retroreflectors or devices of comparable tracking accuracy to the lunar surface, it would be possible to improve the lunar science results and tests of gravitational physics.  Lunar ranging should continue to investigate the lunar interior and probe the nature of gravity with improved accuracy.

\section*{Acknowledgments}
We acknowledge and thank the staffs of the Observatoire de la Côte d'Azur, Haleakala, and University of Texas McDonald ranging stations.  A. Konopliv provided the radio range and Doppler accuracies.  The research described in this paper was carried out at the Jet Propulsion Laboratory of the California Institute of Technology, under a contract with the National Aeronautics and Space Administration.


\end{document}